**Comment on „Microscopic modeling of multi-lane highway traffic flow", Nathan O. Hodas and Arnand Jagota, Am. J. Phys. 71 (12) 2003, pp. 1247**

M. Risch, University of Applied Sciences, Augsburg, Germany

In heavy traffic with congested roadway the maximum traffic flow (road capacity) also depends on length of cars (minimum clearances). This is deduced in a simple derivation suited for classroom demonstration as well as homework. The resulting equation demonstrates a new relation to an apparently unrelated area of physics, the maximum ship velocity (hull speed) and explains why traffic is sometimes faster on the slow (right) lane on a congested multi-lane road.

Capacity of a road is determined by car velocity and distance between cars, besides other influences. Maximum capacity occurs at a specific speed, it usually is attained by self-regulating on a congested road. This equilibrium velocity can be derived using simplifying assumptions by derivation of an equation.

In a congested roadway capacity q of the road is determined by velocity v and following distance s of cars; v and s depend on each other. Following distance s of cars is a fixed fraction of the way to stop the car, typically 50% to 100 % [1,2]. For sake of simplicity this fraction is set here to 1 (100%) without affecting essentials of results. Then s becomes sum of breaking distance and distance driven during reaction time $t_h$,

$$s = \frac{v^2}{2g} + v \cdot t_h$$

It is assumed that maximum break acceleration is gravity g (friction coefficient µ times gravity, µ having a maximum of 1, stronger breaking would lock wheels[3,4]). Calculation of the fraction of road length consumed by the car L/N requires to add the cars length l summing to

$$\frac{L}{N} = \frac{v^2}{2g} + v \cdot t_h + l$$

This neglects influences of slope and curvature of roads which practically often limit capacity [5]. Capacity of a road $q_{max}$ is maximum traffic flow q equaling number of cars

past a region of a lane in time t

$$q = \frac{v}{L/N} = \frac{2g \cdot v}{v^2 + 2g \cdot v \cdot t_h + 2g \cdot l}$$

This function q demonstrates following dependencies:

a) At v and l constant q continuously decreases with $t_h$

b) At v and $t_h$ constant q continuously decreases with l

c) At $t_h$ and l constant q has a maximum at $v_{max}$ resulting in a $q_{max}$

For example:

At l=5m and $t_h$ = 1s $v_{max}$ is 10 m/s = 36 km/h = 22mi/h and maximum traffic flow $q_{max}$ is 1700 cars/h/lane in agreement with observations [1,2,5].

$q_{max}$ occurring at $v_{max}$ is usually attained by self-regulating on a congested road [2,3,4].

$q_{max}$ is determined by the derivative of function q (v):

$$q' = \frac{dq}{dv} = 2g \cdot \frac{2g \cdot l - v^2}{(v^2 + 2g \cdot v \cdot t_h + 2g \cdot l)^2}$$

Maximum traffic flow $q_{max}$ is attained when

$$2g \cdot l - v^2 = 0$$

yielding to a square root function

$$v_{max} = \sqrt{2g \cdot l} \qquad (1)$$

Examples: car l = 5m $v_{max}$ = 10 m/s = 36 km/h, Truck l =15 m $v_{max}$ = 17 m/s = 60 km/h

In traffic on a congested road, the slow (right) lane usually has trucks with longer l resulting in faster equilibrium speeds first, which will eventually even out by drivers choosing to change lanes.

Result (1) most curiously demonstrates a new relation to an apparently unrelated area of physics, the maximum ship velocity $v_{mahu}$ (hull speed) of a ship displacing water and floating in deep water (deeper than ship length l) which is due to water wave Dispersion and resonance

$$v_{mahull} = \sqrt{g \cdot l / 2\pi} \qquad \text{provided} \qquad l < H \text{ (Water- depth)}$$

This new relation to an apparently unrelated area of physics might show students physics can be of great simplicity.

email    mrrisch@rz.fh-augsburg.de